\documentclass[preprint2,longabstract]{aastex}

\usepackage{graphicx}
\usepackage{amssymb}
\usepackage{natbib}
\tabletypesize{\normalsize}  

\def\plotone#1{\centering \leavevmode
\includegraphics[width=.95\columnwidth]{#1}}

\shortauthors{Wallerstein et al.}
\shorttitle{Metal Abundance Calibration of the Ca~{\sc ii} Triplet}
\begin{document}
\large
\pagenumbering{arabic}

\title{Metal Abundance Calibration of the Ca~{\sc ii} Triplet Lines in RR Lyrae Stars}

\author{{\noindent George Wallerstein\altaffilmark{1}, Thomas Gomez\altaffilmark{1,2}, and Wenjin Huang\altaffilmark{1,3}\\{\it (1) Department of Astronomy, University of Washington, Seattle, WA 98195} 
}
}
\email{(1) wall@astro.washington.edu}

\altaffiltext{2}{Present address: Department of Astronomy, University of Texas at Austin; gomezt@astro.as.utexas.edu}
\altaffiltext{3}{Present address: Brion, Inc.; wenjin.huang@brion.com}

\begin{abstract}
The Gaia satellite is likely to observe thousands of RR Lyrae Stars with a small spectral range, between 8470\AA~to 8750\AA, at a resolution of 11,500. In order to derive metallicity from Gaia, we have obtained numerous spectra of RR Lyrae stars at a resolution of 35,000 with the Apache Point Observatory 3.5m echelle spectrograph. We have correlated Ca~{\sc ii} triplet line strengths with metallicity as derived from Fe~{\sc ii} abundances, analogous to Preston 1959, ApJ, 130, 507 use of the Ca~{\sc ii} K line to estimate metallicity in RR Lyrae Stars. 

\end{abstract}


\section{Introduction: the Gaia Satellite}

During the next decade, the study of galactic structure will be greatly enhanced thanks to two new instruments, the Large Synoptic Survey Telescope (LSST) and the Gaia satellite. The European Space Agency (ESA) is heading the Gaia satellite project, an enormous improvement over {\it Hipparcos}. 

\begin{figure}
\begin{center}
\epsscale{1.0}
\plotone{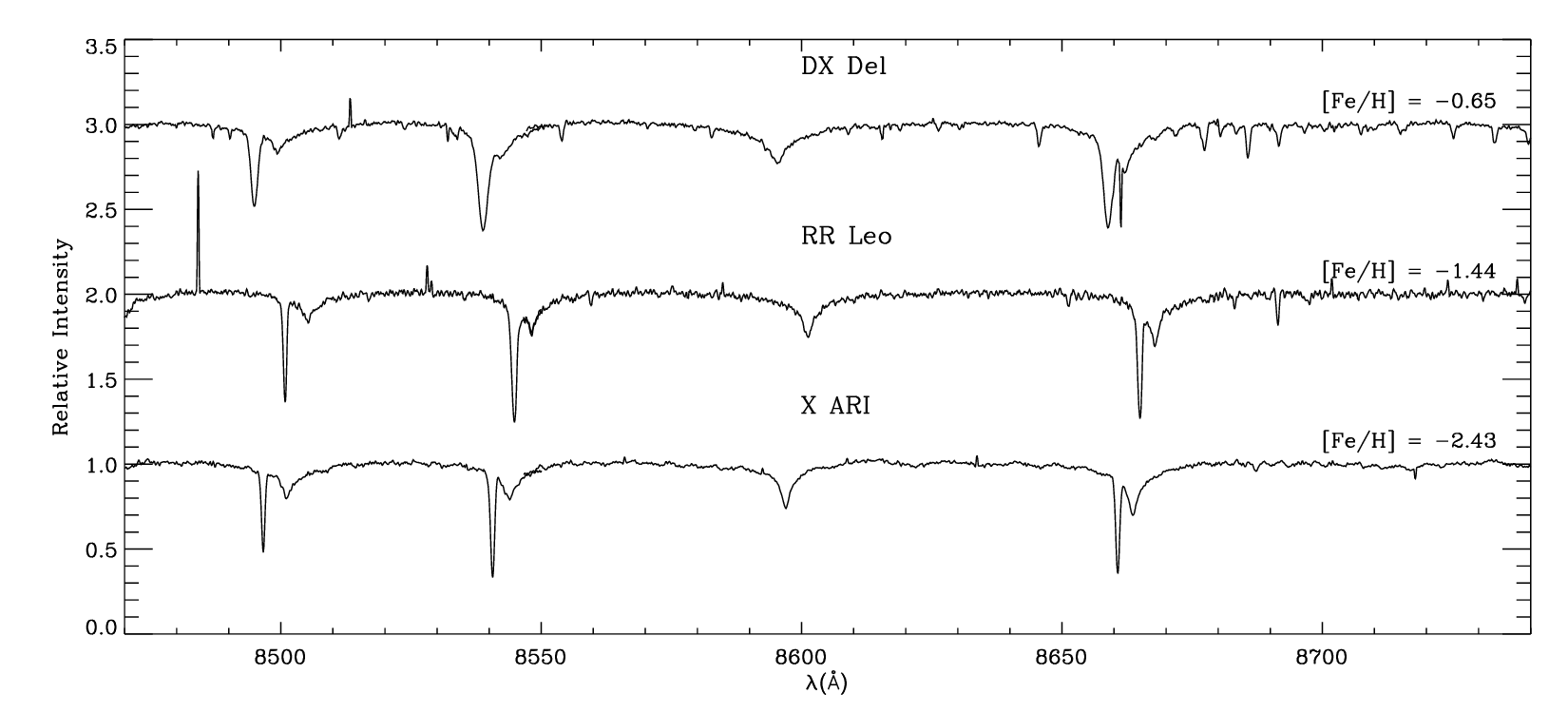}
\caption{This demostrates the full spectral range of Gaia at Gaia resolution. All the stars in the presented spectra have an effective temperature of about 6500K, this is to demonstrate the spectra at different metallicities. Notice how the Ca~{\sc ii} triplet lines become more blended with their neighboring Paschen lines.}
\label{gaia}
\end{center}
\end{figure}

In addition to measuring parallaxes and proper motions, Gaia is expected to measure a small spectral window between 8470\AA~and 8750\AA~ with a resolution of 11,500 to capture the Ca~{\sc ii} triplet for radial velocity measurements. This expected spectral window also includes some of the Paschen hydrogen series. Gaia's mission plan is to measure each star multiple times in order to get other information about each star, for example, if the star is variable and to better refine various measurements. The spectra taken over the course of the mission will be averaged to get a radial velocity. The Gaia spectral window is shown for several RR Lyrae (X Ari, RR Leo and DX Del) in Figure \ref{gaia}. The spectra presented have been reduced from the Apache Point Observatory (APO) resolution to the expected Gaia resolution. We correlated [Fe/H] against the EW of the Ca~{\sc ii} triplet for RR Lyrae in the same manner as Cole et al. (2004) did for red giants. This is important because Gaia will map an estimated one billion stars, including many RR Lyrae variable stars, which are important for galactic structure and stellar populations. This relationship allows us to know one of the fundamental properties of RR Lyrae without additional information.

\section{Observations and Data Analysis}

We have observed many RR Lyrae stars with the APO echelle spectrograph, which has a resolution of 35,000, that covers a wavelength interval from 3500\AA~to 10,400\AA~ with good signal-to-noise in the 8500\AA~region (S/N ranged from 70 to 200). Data reduction was accomplished using programs in the IRAF\footnote{IRAF is distributed by the National Optical Astronomy Observatories, which are operated by the Association of Universities for Research in Astronomy, Inc., under cooperative agreement with the National Science Foundation} package. Equivalent width measurements were obtained both at the full resolution of APO's echelle and the expected reduced resolution of Gaia at 11,500. Spectra of the RR Lyrae stars were observed at random phases in order to replicate Gaia observations. 

The measurement of equivalent widths (EW) was automated by a program written by WH in IDL. Table 1 shows a list of target stars observed at APO and their properties as well as observation dates. Lines of both the hydrogen Paschen and Balmer series, Ca~{\sc ii} triplet, and selected lines of Fe~{\sc ii}, Fe~{\sc i}, and other elements were measured. The EWs of Fe~{\sc ii} were analyzed with a Kurucz (1993) atmosphere model to determine [Fe/H].  Further abundance analysis of these other elements will be presented in the full publication. 

\begin{figure}
\epsscale{1.0}
\plotone{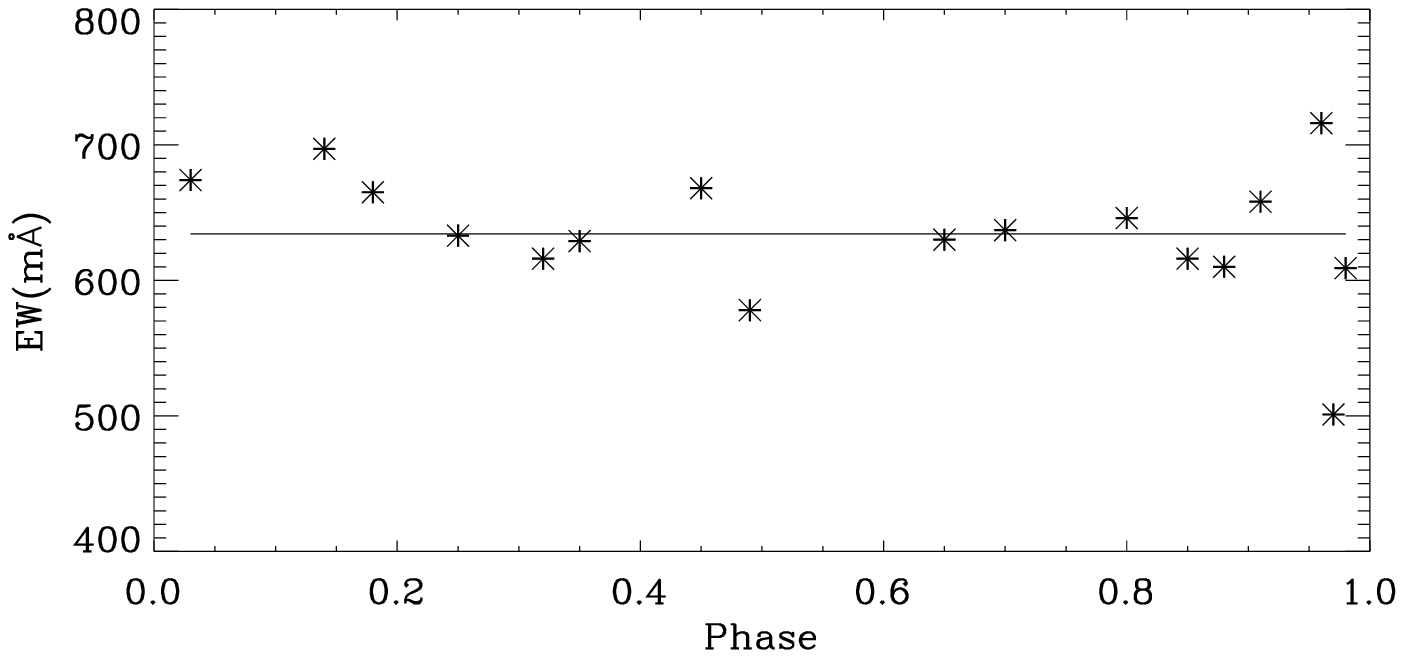}
\caption{Continuous observations of AS Vir allow us to study the variations of the Ca~{\sc ii} triplet line at 8498\AA. The range of data shows that close to maximum temperature the profile becomes difficult to accurately measure, hence the scatter in the plot. The average EW is 634.3m\AA with a standard deviation of 48.64m\AA. The ratio of the standard deviation to the average is 7.6\%. Data provided by Chris Sneden, George Preston, and BiQing For}
\label{phase}
\end{figure}

A collaboration with the Chris Sneden, BiQing For, and George Preston has given us access to continuous observations of some southern hemisphere RR Lyrae not accessible from APO. Figure \ref{phase} demonstrates the stability of the Ca~{\sc ii} for use in our analysis. Around the beginning of the pulsation cycle the profiles show emission features at the core, making it difficult to measure. Since our observations at APO simulate the expected observing cadence of Gaia, very few observations that Gaia will make of RR Lyrae will catch this phase of pulsation.

\begin{deluxetable}{lcc|lcc}
\tabletypesize{\normalsize}
\tablecolumns{6}
\tablecaption{Lists of observed stars (sorted by constellation), their [Fe/H] and [Fe/H] reference. The observation date and time are included.}
\tablehead{
  \colhead{Identifier}& 
  \colhead{[Fe/H]}&
  \colhead{UT at Observation}
 }
\startdata
AT And & $-$1.18\tablenotemark{a} & 2010-11-21 04:22 & RX Eri & $-$1.33\tablenotemark{b} & 2010-11-21 06:42\\
SW And & $-$0.38\tablenotemark{a} & 2009-12-27 05:01 & SV Eri & $-$2.04\tablenotemark{a} & 2009-12-27 06:17\\
XX And & $-$1.94\tablenotemark{b} & 2010-11-21 05:14 & & & 2009-11-27 06:00\\
BR Aqr & $-$0.74\tablenotemark{b} & 2010-11-21 03:50 & RR Gem & $-$0.35\tablenotemark{a} & 2009-12-27 06:45\\
BS Aqr & & 2008-12-16 00:53 & & & 2008-12-16 06:03\\
X Ari  & $-$2.43\tablenotemark{b} & 2010-11-21 05:40 & VX Her & $-$1.58\tablenotemark{b} & 2009-03-15 10:50\\
& & 2009-12-27 05:52 & VZ Her & $-$1.03\tablenotemark{a} & 2010-03-25 10:38\\
& & 2009-11-27 05:35 & & & 2009-03-15 11:34\\
& & 2009-08-04 11:33 & RR Leo & $-$1.57\tablenotemark{a} & 2009-03-15 08:21\\
RS Boo & $-$0.32\tablenotemark{a} & 2010-03-25 07:17 & TT Lyn & $-$1.56\tablenotemark{b} & 2009-12-27 08:09\\
& & 2009-06-08 06:02 & TT Lyn & $-$1.56\tablenotemark{b} & 2009-12-27 08:09\\
ST Boo & $-$1.86\tablenotemark{a} & 2009-03-15 09:25 & & & 2009-03-15 08:09\\
RR Cet & $-$1.52\tablenotemark{a} & 2009-11-27 03:45 & KX Lyr & $-$0.46\tablenotemark{a} & 2010-03-25 11:13\\
& & 2009-08-04 10:30 & & & 2009-06-08 06:29\\
XZ Cet &  & 2009-11-27 04:39 & RR Lyr & $-$1.37\tablenotemark{a} & 2009-12-27 00:55\\
W CVn  & $-$1.21\tablenotemark{a} & 2009-06-08 04:58 & & & 2009-08-04 09:41\\
& & 2009-03-15 09:05 & & & 2009-06-13 07:32\\
XZ Cyg & $-$1.52\tablenotemark{a} & 2009-08-04 09:57 & & & 2009-03-15 12:22\\
& & 2009-06-13 07:49 & V445 Oph &-0.23\tablenotemark{a} & 2009-06-08 05:26\\
& & 2008-12-16 01:36 & & & 2009-03-15 11:11\\
SW Dra & $-$1.24\tablenotemark{a} & 2010-03-25 08:27 & AV Peg & $-$0.14\tablenotemark{a} & 2009-06-13 08:58\\
& & 2009-05-09 08:26 & BH Peg & $-$1.38\tablenotemark{a} & 2009-12-27 02:36\\
XZ Dra & $-$0.87\tablenotemark{a} & 2010-03-25 11:50 & AR Per & $-$0.43\tablenotemark{a} & 2009-12-27 07:42\\
& & 2009-12-27 01:11 & TU UMa & $-$1.44\tablenotemark{b} & 2009-03-15 10:12\\
& & 2009-06-13 08:06 & UU Vir & $-$0.82\tablenotemark{a} & 2010-03-25 07:59\\
DX Del & $-$0.56\tablenotemark{a} & 2009-12-27 01:36 & & & 2009-03-15 09:50\\
& & 2009-11-27 01:46\\
& & 2009-08-04 09:16\\
& & 2009-06-13 08:33\\
& & 2008-12-16 01:22\\
\enddata
\label{prop}
\tablenotetext{a}{\it{Layden (1994)}}
\tablenotetext{b}{\it{Feast et al. (2008)}}
\end{deluxetable}

\begin{figure}
\begin{center}
\epsscale{1.0}
\plotone{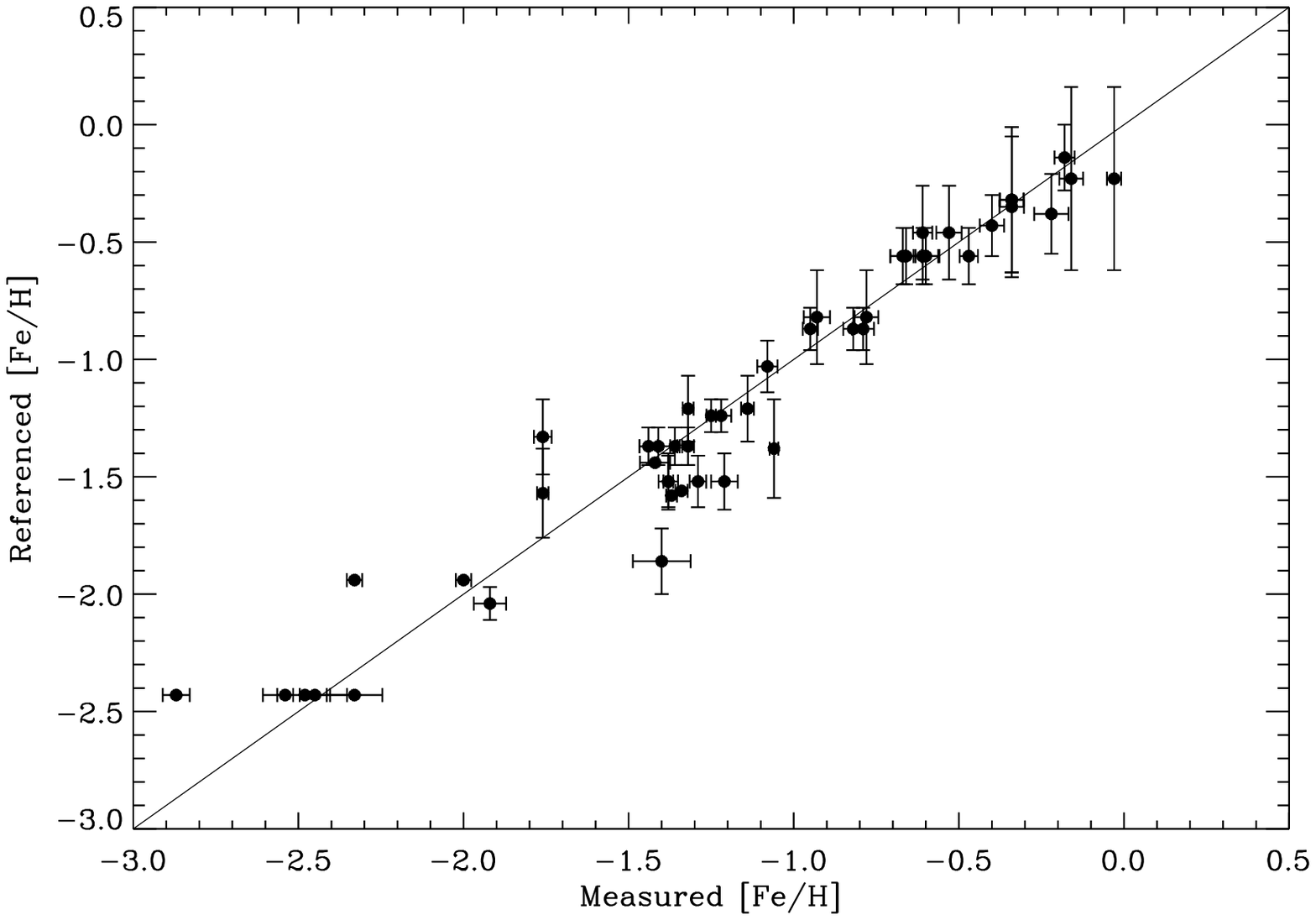}
\caption{The derived [Fe/H] compared to the references, given in Table \ref{prop}. We show individual measurements on this plot, for example, we have 5 X Ari measurements corresponding to the points at [Fe/H] = -2.5.}
\label{FeH}
\end{center}
\end{figure}

The temperature of each star at the observed phase was determined from the hydrogen Balmer and Paschen lines using Kurucz models. Based on the known luminosity of RR Lyrae stars and an estimated mass of 0.7 $M_{\odot}$, we assume log $g=2.5$ unless the star is hotter than 6750~K. Stars with $T_{\rm eff}$ $\geq$ 6750~K are assumed to have log $g=3.5$. The appropriate Kurucz model was then used in MOOG to determine the abundances of other elements. Our Fe abundances compared to values from the literature are shown in Figure \ref{FeH}.

\section{Metallicity Relation}

Preston (1959) was able to relate [Fe/H] with $\Delta$S, the difference between spectral types derived from the Ca~{\sc ii} K line and the Balmer series. Clementini et al. (1991) were able to derive a tighter correlation between [Fe/H] and the Ca~{\sc ii} K EW. We believe that the Ca~{\sc ii} triplet rather than H and K will provide a comparable (if not better) metallicty relationship due to the triplet's unblended and sharp profiles.

\begin{figure}
\begin{center}
\epsscale{1.5}
\plotone{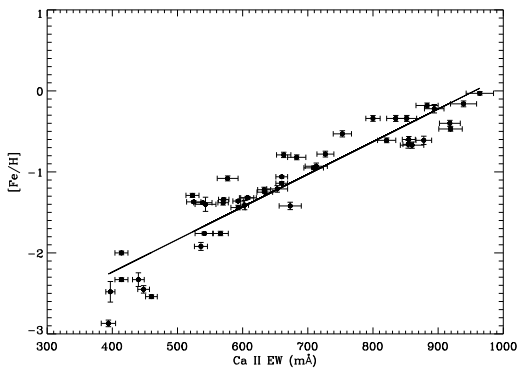}
\caption{The relationship between the Ca~{\sc ii} at 8498\AA~to our derived [Fe/H]. The least squares fit is best described by equation 1. Here we again show individual measurements of [Fe/H] and the Ca~{\sc ii} triplet.}
\label{CaII}
\end{center}
\end{figure}

To derive a similar relationship between [Fe/H] and the EW of the Ca~{\sc ii} infrared triplet, the line that can best be measured is the Ca II line at 8498\AA~ because it can be fitted with a Gaussian and is only minimally blended with the Paschen line at 8502\AA, see Fig \ref{gaia}. The small separation between the other Ca II lines and their neighboring Paschen lines makes measuring them more difficult, especially for metal rich stars. Measurements are best made when the Paschen lines are at their weakest in order to best determine the EW. The comparison of [Fe/H] to the EW of the 8498\AA~Ca~{\sc ii} line of each star is shown in Figure \ref{CaII}. 

In conclusion we show that there is a relationship between [Fe/H] and the EWs of the Ca~{\sc ii} infrared triplet given by

\begin{equation}
[Fe/H]~=~-3.846(\pm0.155) + W*0.004(\pm0.0002)
\end{equation}

\subsection{The Ca/Fe relationship}
In response to a question raised by Rolf-Peter Kudritzki in the subsequent discussion of our presentation we have employed lines of Ca~{\sc i} and Fe~{\sc i} to derive preliminary ratios of [Ca/Fe] based on the LTE population ratios of neutral to ionized species. Since the fraction of neutral atoms at the temperatures of RR Lyrae stars is miniscule, an nLTE formulation would be preferable so we refer to our ratios as preliminary.

\begin{figure}
\epsscale{1.0}
\plotone{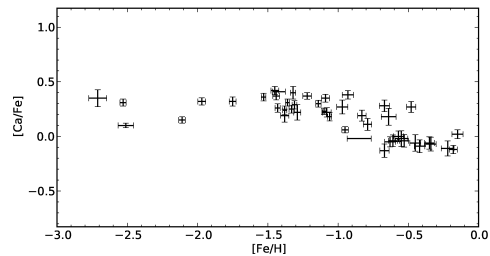}
\caption{The [Ca~{\sc i}/Fe~{\sc i}] abundance as a function of the Fe~{\sc ii} abundance of the RR Lyrae. Each measurement is shown in the plot.}
\label{CaFe}
\end{figure}

We have selected lines with a wide range of gf-values and excitations because they are extremely weak in metal-poor stars near maximum light and very strong near minimum for stars with [Fe/H] near zero. Due to space limitations full details will be published elsewhere. We have used f-values tabulated by VALD and derived abundances from the measured EW via MOOG using the atmosphere models by Kurucz. We show the logarithmic ratio, [Ca/Fe], as a function of [Fe/H] in Figure \ref{CaFe}. Figure \ref{CaFe} shows every measurement taken, even for stars where more than one measurement was taken. The correlation shown in Figure \ref{CaFe} is in agreement with the relationship of the ratio of the alpha-elements to Fe first recognized by Wallerstein (1962) and confirmed by many subsequent studies. [Ca/Fe] rises from near zero for the most metal-rich stars to about [Ca/Fe] = 0.4 near [Fe/H] = -1.0 and remains constant down to [Fe/H] = -2.5.

\acknowledgements

We thank Ulisse Munari and Gisella Clementini for advice and assistance. This research was supported by the Kennilworth fund of the New York Community Trust. We would like to thank George Preston, Chris Sneden and BiQing For for their data of AS Vir. We also thank the UW McNair program for their support of T. Gomez.


\end{document}